\documentclass[11pt,a4paper]{article}
\usepackage{amsmath,amsfonts,amssymb,amscd}

\newenvironment{Proof}{\par\noindent{\bf Proof}\,}{\hskip.4em$\Box$\par\smallskip}
\newtheorem{Theorem}{Theorem}
\newtheorem{Definition}[Theorem]{Definition}
\newtheorem{Proposition}[Theorem]{Proposition}

\def\plus#1#2{\vrule height#1pt width0pt depth#2pt}
\def\@{\hskip.8pt}
\def\?{\hskip.3pt}

\def\Ref#1#2{\if#2)\ref{#1}#2\else\ref{#1}\@#2\fi}

\def\h#1{\hat{#1}}

\def\Eps#1{\eps_{\lower1pt\hbox{$\scriptstyle #1$}}}

\def\d#1/d#2{\frac{d\/#1}{d\/#2}}
\def\de#1/de#2{\frac{\partial\/#1}{\partial\/#2}}
\def\SD#1/de#2/de#3{\ifx#2 \frac{\plus02\partial^{\@\@2}#1}
    {\plus90\partial\@#3^{\@2}} \else\frac{\plus02\partial^{\@\@2}#1}
    {\partial\?#2\partial\?#3}\fi}
\def\TD#1/de#2/de#3/de#4{\frac{\plus02\partial^{\@\@3}#1}%
    {\partial\?#2\partial\?#3\partial\?#4}}
\def\D#1/D#2{\frac{D\/#1}{D\/#2}}

\def\DD#1/D#2{\textstyle{\text{\Large$\D{#1}/D{#2}$}}}
\def\dd#1/d#2{\textstyle{\text{\Large$\d{#1}/d{#2}$}}}
\def\De#1/de#2{\textstyle{\text{\Large$\de{#1}/de{#2}$}}}
\def\sD#1/de#2/de#3{\textstyle{\text{\Large$\SD{#1}/de{#2}/de{#3}$}}}

\def\iff{\Leftrightarrow}

\begin{document}
\vskip-2cm

\title{Existence of solutions for Hamiltonian field theories by the Hamilton-Jacobi technique.}
\author{Danilo Bruno \\
        Dipartimento di Matematica dell'Universit\`a di Genova \\
        Via Dodecaneso, 35 - 16146 Genova (Italia) \\
        E-mail: bruno@dima.unige.it
          }
\date{}
\maketitle

\begin{abstract}
\noindent
The paper is devoted to prove the existence of a local solution of the Hamilton-Jacobi equation in field theory, whence the general solution of the field equations can be obtained. The solution is adapted to the choice of the submanifold where the initial data of the field equations are assigned. Finally, a technique to obtain the general solution of the field equations, starting from the given initial manifold, is deduced. 
\par\bigskip
\noindent
{\bf 2010 Mathematical subject classification:} 70H20, 70S05, 35F21
\newline
{\bf Keywords:} Hamiltonian field theories, Hamilton-Jacobi equation. 
\end{abstract}
\thispagestyle{empty}

\section{Introduction}

Hamilton-Jacobi theory plays an important role within classical point particle mechanics, being a powerful tool  to solve the evolution equations on the one hand, and the cornerstone for the transition from classical to non--relativistic quantum mechanics in Schroedinger' s formulation on the other. 

The formulation of Hamilton--Jacobi equation for field theories is quite old  and possesses many different equivalent approaches\cite{Rund}. It has been recently revised for its relationship with relativistic quantum mechanics\cite{Rovelli} and for its geometric properties\cite{Leon1}.  

Like the classical case, the interest of the Hamilton-Jacobi method is placed in the possibility of using it to obtain solutions of the field equations. Indeed, the possibility of using the concept of complete integral to perform this operation is precluded, being its existence quite an exceptional case in field theory. 

In fact, while the equation itself is even less constraining than the mechanical one, being a single PDE in many unknown functions, an additional set of {\it embeddability conditions} is needed to relate its solutions to the field equations. This results into a set of quite restricting conditions and the only available possibilities of finding solutions  strongly depend on the nature of the particular problem under consideration. 

Many different authors pursued the aim of producing sets of particular solutions: some of them built single solutions starting from  given solutions of the field equations\cite{vonRieth,Rovelli}, others use the concept of Baecklund transformation under particular dimensional restrictions\cite{Kastrup}. In a previous paper\cite{b1}, a particular set of solutions was constructed; they were not constrained by dimensionality requirements, but the embeddability conditions resulted into a set of quite restricting compatibility relations on the initial data. 

This paper is devoted to proving the existence of  solutions of the Hamilton--Jacobi equation in field theory, that allow to determine the general solution of the field equations. The only restrictions are placed in a set of quite general regularity conditions on the submanifold where the initial data are assigned. For this purpose, an algorithm is constructed, starting from a particular solution of the Hamilton--Jacobi equation, adapted to the choice of the initial manifold: this approach differs from the previous ones in its being sufficiently general to avoid that the embeddability conditions impose any restriction on the possible solutions of the field equations. 

Although the general solution was obtained, a complete integral, generating the whole parametric set of solutions, similar to the particle mechanics case cannot be constructed, being the solution still dependent on the choice of the initial surface. The solution of the field equation by means of the Hamilton-Jacobi method can however be obtained using an alternative concept of  complete integral, adapted to the initial submanifold. This method is still a very partial algorithm, since it is not able to automatically include the embeddability conditions, that need to be integrated apart and is a current argument of research. \\[4pt]
\indent The paper is organized as follows:\\
Section II is devoted to recalling the main features of Hamiltonian field theory, which represents the background where Hamilton-Jacobi equation is placed. In particular the link between the solution of the field equations and the Hamilton-Jacobi theory is strictly revised. All the arguments are already contained in \cite{b1}. \\
Section III describes the new technique for solving the Hamilton-Jacobi equation. 
The procedure builds a solution of the Hamilton-Jacobi equation, whence the solution of the field equations is deduced, for a general enough choice of the initial data. The proof of the existence of such a solution is limited to the class of analytic functions, being Cauchy-Kovalevskaya theorem needed to complete the argument. 
Section IV is devoted to showing possible future developments of the argument in the direction of defining  a new notion of complete integral, adapted to the initial submanifold. \\
In section V a simple example is given, in order to show how the whole machinery works. 

\section{Hamiltonian Field theory and the Hamilton-Jacobi equation}

The Hamilton--Jacobi theory represents a tecnique for solving the differential equations of field theory in their Hamiltonian description. In this section, the most relevant aspects of Hamiltonian field theory will be briefly revised. The reader is referred to \cite{b1} for the details. The paper will directly deal with field equations in their Hamiltonian version; the relationship with the Lagrangian counterpart will be assumed as known.\\[2pt]
\noindent
$\bullet$ Field theories are generally described from a geometric viewpoint taking a fiber bundle $\pi: E\to M$ into account, endowed with a set of local coordinates $x^1,\ldots,x^n,y^1,\ldots,y^r$, the first set denoting the basis coordinates on $M$ and the remaining ones the local coordinates on the fibers. Within this context, a physical field is then represented as a section $\varphi : M\to E$, locally written in the form:
\[
y^i = \varphi^i\/(x^\mu) \quad i=1,\ldots,r \; , \; \mu = 1,\ldots,n \; .
\]
The base manifold $M$ is supposed to be endowed with a (pseudo)-metric tensor $\Phi$. \\[2pt]
$\bullet$ Hamiltonian field theory is developed within the framework provided by the bundle $\Lambda^n\/(E)$ of skew--simmetric $n$-forms on $E$ (compare with \cite{cvb4}). Let us take the bundle $\Lambda^n_1\/(E)$ of horizontal $n$-forms into account, whose elements are annihilated whenever one of its arguments is vertical. Then, let $\Lambda^n_2\/(E)$ be the bundle of $1$-contact forms on $E$, whose elements have the property of vanishing when at least two of their arguments are vertical vector fields. 
The elements of the two bundles can be locally described as follows:
\begin{subequations}\label{2.1}
\begin{equation}
\sigma \in \Lambda^n_1\/(E) \quad \iff \quad \sigma = p_0\/(\sigma)\,ds \quad , 
\end{equation}
\begin{equation}
\sigma \in \Lambda^n_2\/(E) \quad \iff \quad \sigma = p\/(\sigma)\,ds + p^\mu_i\/(\sigma)\,dy^i\wedge ds_\mu \quad . 
\end{equation}
\end{subequations}
Then, the bundles $\Lambda^n_1\/(E)$ and $\Lambda^n_2\/(E)$ are respectively described by a system of local coordinates $x^\mu,y^i,p_0$ and $x^\mu,y^i,p,p^\mu_i$; moreover, 
$\Lambda^n_1\/(E)\to E$ is a vector sub--bundle of  $\Lambda^n_2\/(E) \to E$.  The quotient bundle $\Pi\/(E):=\Lambda^n_2\/(E) / \Lambda^n_1\/(E)$ will be henceforth called the {\it phase space} and endowed with a set of local coordinates $x^\mu,y^i,p^\mu_i$. It is easy to prove that $\Lambda^n_2\/(E)\to \Pi\/(E)$ is an affine bundle, modelled on $\Lambda^n_1\/(E)$. \\[2pt]
$\bullet$ The assignment of the dynamical information is summarized giving a section ${\cal S} : p = -H\/(x^\mu,y^i,p^\mu_i)$ of the bundle $\Lambda^n_2\/(E)\to \Pi\/(E)$; the latter is called the {\em Hamiltonian section} and the function $H$ is called the Hamiltonian function. It can be obtained from the knowledge of the Lagrangian of the theory through the Legendre transformation. The field equations, obtained from a variational problem on the phase space, can be written in terms of the Hamiltonian as follows:
\begin{equation}\label{2.2}
\d y^i /d{x^\mu} = \de H /de{p^\mu_i} \quad ; \quad \d p^\mu_i/d{x^\mu} = - \de H /de{y^i}\quad .
\end{equation}
The 	possibility of deducing eqs.\eqref{2.2} from a Lagrangian formulation is related to the {\em regularity condition}
\begin{equation}\label{regularity}
\det \left(\SD H/de{p^\mu_i}/de{p^\nu_j}\right) \neq 0 \quad ,
\end{equation}
that will be tacitly assumed from now on. The calculation of the above determinant can be performed choosing an ordering, mapping the pair $(\mu,i)$ into a single index, running from 1 to $n\times r$; the vanishing of the result is independent on the choice of the order. 

\noindent
The field equations have the nature of a system of first-order partial differential equations for the functions $y^i = y^i(x^\mu), p^\mu_i = p^\mu_i(x^\mu)$. They are generally given together with a set of initial conditions, which  essentially consist in the assignment of the fields and of their transverse derivatives over an $(n-1)$-dimensional submanifold $i:N\to M$, locally expressed as 
\[
i: x^\mu = \varphi^\mu\/(z^1,\ldots,z^{n-1}) \quad , \quad \mu = 1\ldots n \quad , 
\]
where $z^1,\ldots,z^{n-1}$ represent a system of local coordinates on $N$. The whole set of initial data is therefore represented as:
\begin{equation}\label{Dati iniziali}
\left\{
\begin{aligned}
&x^\mu = \varphi^\mu\/(z^A)\quad &&  A=1\ldots n-1 \\
&y^i = \psi^i\/(z^A)\quad  &&  \mu = 1\ldots n \\
&\de y^i/de{x^\mu} n^\mu = \h\psi^i\/(z^A)\quad  && i=1\ldots r
\end{aligned}
\right.\quad , 
\end{equation}
where $n = n^\mu \de /de{x^\mu}$ is a vector field defined on $N$ and transverse to it.
Whenever the regularity condition \eqref{regularity} is satisfied, the equation
\[
y^i_\mu = \de H/de{p^\mu_i} (x^\mu,y^i,p^\mu_i)
\]
can be locally inverted with respect to $p^\mu_i$. This allows to write eq.\eqref{Dati iniziali} in the form
\begin{equation}\label{Superficie dati iniziali}
{\cal B} : 
\left\{
\begin{split}
&x^\mu = \varphi^\mu\/(z^A) \\
& y^i = \psi^i\/(z^A)\\
& p^\mu_i = \tilde\psi^\mu_i(z^A) \\
& p = -H(\varphi(z^A),\psi(z^A),\tilde\psi^\mu_i(z^A))
\end{split}
\right..
\end{equation}
representing a $(n-1)$-dimensional  submanifold of $\Lambda^n_2(E)$ \\[2pt]
\noindent
$\bullet$ The Hamilton-Jacobi equation can be obtained determining a particular foliation of the Hamiltonian section  ${\cal S}$. For this purpose, let $\sigma\in \Lambda^{n-1}_1\/(E)$ be a generic horizontal $(n-1)$-form, locally described as:
\begin{equation}\label{3.1}
\sigma = S^\alpha\/(x^\mu,y^i)\,ds_\alpha\quad . 
\end{equation}
Its differential $\sigma\in \Lambda^{n}_2\/(E)$ is locally described by
\begin{equation}\label{3.2}
d\sigma = \de S^\alpha\/(x^\mu,y^i)/de {x^\alpha} ds + \de S^\alpha\/(x^\mu,y^i)/de{y^i} dy^i\wedge ds_\alpha \quad . 
\end{equation}
and represents a a section $\Sigma:E\to\Lambda^{n}_2\/(E)$, locally written as
\begin{equation}\label{3.3}
\Sigma: \quad p =  \de S^\alpha\/(x^\mu,y^i)/de {x^\alpha} \quad , \quad p^\alpha_i =  \de S^\alpha\/(x^\mu,y^i)/de{y^i}  \quad . 
\end{equation}
The latter is contained in the Hamiltonian section if the following condition is satisfied:
\begin{equation}\label{HJ}
\de S^\alpha/de{x^\alpha} + H\/\left(x^\mu,y^i,\de S^\alpha/de{y^i}\right) = 0 \quad . 
\end{equation}
Eq.~\eqref{HJ} is known as the Hamilton-Jacobi equation for field theories. \\[2pt]
$\bullet$ The solutions of the Hamilton-Jacobi equation are related to the solutions of field equations. Taking the derivative of \eqref{HJ} by $y^i$ we find that
\[
\de /de{x^\alpha} \de S^\alpha /de{y^i} + \de H /de{y^i} + \SD S^\alpha /de y^i /de {y^j} \de H /de {p^\alpha_j} = 0 \;  \Rightarrow \; \de /de{x^\alpha} \left( \de S^\alpha /de{y^i}\/(x^\mu,y^i\/(x^\mu)) \right)  = - \de H /de{y^i} \quad . 
\]
This means that the quantities $p^\alpha_i = \de S^\alpha/de{y^i}$ fulfill a subset of the field equations, for any choice of the functions $y^i(x^\mu)$. Unluckily, the rest of the field equations are not necessarily fulfilled, and need to be checked by hand on the solution. The resulting condition is called {\em embeddability condition} and is represented as
\begin{equation}\label{embeddability}
\de y^i /de{x^\mu} = \de H /de{p^\mu_i} \left(x^\mu, y^i,\de S^\mu /de {y^i} \right) \quad . 
\end{equation}
A solution $y^i = y^i\/(x^\nu)$, $p_i^\mu = p_i^\mu\/(x^\nu)$ of the field equations is said to be embeddable in a solution $S^\mu = S^\mu\/(x^\nu,y^i)$ of \eqref{HJ} if condition \eqref{embeddability} holds. 

\noindent
If $\mu=1$, i.e. in the case of point particle mechanics, eq.~\eqref{embeddability} describes a velocity field on $E$ for every solution of \eqref{HJ} and its integrability is guaranteed by Cauchy theorem; therefore, it is possible to determine a family of solutions of the field equations embedded in every solution of the Hamilton--Jacobi equation, at least locally.  In the general case, this is not necessarily true; the reader is referred to \cite{b1} for an analysis of the possible scenarios present in the literature.\\[2pt]

\noindent
The approach adopted here will be that of determining a parametric set of solutions of the Hamilton--Jacobi equation, associated to a chosen initial submanifold $N$. This is made possible by the fact that  eq.~\eqref{HJ} has the nature of a single first--order partial differential equation in the $n$ unknown functions $S^\alpha\,(x^\mu,y^i)$ and therefore possesses a great amount of equivalent solutions. This arbitrariness will be eventually fixed by the embeddability condition. The argument extends the analogous procedure presented in \cite{b1}, where a particular choice of the candidate solution allowed to build particular classes of solutions, with the embeddability conditions giving rise to a set of compatibility requests on the initial data. 
This time, however, this choice will prove to be general enough to generate the whole set of solutions of the field equations, whenever the initial submanifold $N$ satisfies quite general regularity conditions. 


\section{The general solution}

Let us consider a field theory, where the Hamiltonian $H$ is given, together with the surface $N\subset M$ where the initial data are assigned. 
Then, let us consider a vector field $X\in D^1(M)$, defined in an open subset $U\subset M$, containing the initial submanifold $N$, and locally written as
\[
X = X^\mu(x^1,\ldots,x^n) \de /de{x^\mu} \quad .
\]
If $X$ is chosen to be transverse to the initial submanifold $N$, its integral curves  are locally determined as
\begin{equation}\label{coord adattate}
\d x^\mu /d\xi = X^\mu(x^\mu) \quad \rightarrow \quad x^\mu = x^\mu(\xi,z^A) \quad , 
\end{equation}
where $\xi$ is a parameter along the curves and $z^A$ are the coordinates on $N$, parameterizing the starting point, corresponding to $\xi=0$. The set $\xi,z^1,\ldots,z^{n-1}$ define a system of local coordinates around $N$, henceforth called {\em adapted coordinate system}, so that the submanifold $N$ is represented by the slice $\xi=0$. 
Let us consider the candidate solution
\begin{equation}\label{scelta soluzione}
S^\mu(x^\mu,y^i) = \varphi(x^\mu,y^i)\, X^\mu+A^\mu_i(x^\mu)\,y^i \quad , 
\end{equation}
where  ${\cal A}_i = A^\mu_i(x^\mu) \de /de{x^\mu}$ is a family of $r$ vector fields on $M$. The functions $A^\mu_i$ are arbitrary and the solution will depend parametrically on them. Substituting into \eqref{scelta soluzione} we obtain
\begin{equation}\label{HJ sostituito}
\de X^\mu /de {x^\mu} \varphi + X^\mu \de \varphi /de{x^\mu} + \de A^\mu_i /de{x^\mu} y^i + H(x^\mu,y^i,\de \varphi/de{y^i} X^\mu + A^\mu_i) = 0  \quad . 
\end{equation}
For every choice of the functions $A^\mu_i$, eq.\eqref{HJ sostituito} can be interpreted as a single partial differential equation for the function $\varphi(\xi,y^i)$. We notice that the coordinates $z^A$ play the role of parameters, since no derivative is performed with respect to them. Moreover, the user-defined vector field $X^\mu$ can be chosen in such a way that $\de X^\mu/de{x^\mu} = 0$. This choice allows to remove the dependence of the PDE on $\varphi$, thus simplifying the solution strategy. On the other hand, the choice is not coordinate independent and makes the following procedure strictly based on the chosen initial coordinates $x^1,\ldots x^n$. 
As a matter of fact, this limitation is not relevant and has to be thought on the same footing as  the technique of separation of variables, that, when possible, is necessarily related to the choice of a suitable coordinate system. 

 Taking eq.\eqref{coord adattate} into account, eq.\eqref{HJ sostituito} can be rewritten in the adapted coordinate system $\xi,z^A$ as follows:
\begin{equation}\label{HJ ridotto}
\de \varphi(\xi,z^A) /de \xi + \de A^\mu_i /de {x^\mu} (\xi,z^A)\, y^i + H (\xi , z^A, y^i , \de \varphi /de{y^i} X^\mu + A^\mu_i ) = 0 \quad .
\end{equation}
Eq.\eqref{HJ ridotto} is very similar to an ordinary Hamilton-Jacobi equation, where the role of the Hamiltonian is played by the function $H + \de A^\mu_i/de{x^\mu}\,y^i$. 


In order to prove the existence of a solution and to determine the links with the solutions of the field equations, eq.\eqref{HJ ridotto} will now be analyzed though the method of bicharacteristics (cmp. with \cite{Choquet}). For this purpose, given a single differential equation in a single unknown, we need to rewrite it as an algebraic expression of the form
\[
F(x^i,u,p_i) = 0 \quad , 
\]
where $x^i$ are the independent variables, $u$ represents the unknown function and $p_i \approx \de u/de{x^i}$. The solution is obtained integrating the ordinary differential equations describing the bicharacteristic curves, locally written as
\begin{equation}\label{bicaratteristiche}
\d x^i /d \xi = \de F /de{p_i} \quad , \quad \d p_i /d \xi = - \de F/de {x^i} - \de F /de u p_i\quad , \quad \d u /d \xi = \de F /de u + \de F/de{p_i} p_i \quad , 
\end{equation}
and dragging the initial data along them. 
In the present case, taking the identifications $x^i \sim (\xi,y^i)$, $u \sim \varphi$ and $p_i \sim (u_0,u_i)$ into account, eq.~\eqref{HJ ridotto} can be rewritten as:
\begin{equation}\label{HJ jet}
u_0 + \de A^\mu_i /de {x^\mu} y^i + H(\xi, z^A, y^i , u_i X^\mu + A^\mu_i ) = 0 \quad . 
\end{equation}
Since eq.\eqref{HJ jet} depends {\em parametrically} on the $z^A$s, the solution will be represented as a parametric family of functions. 
The differential equations for the bicharacteristics of eq.\eqref{HJ jet} take the form
\begin{subequations}\label{HJ bicaratteristiche}
\begin{equation}
\d y^i /d\xi = \de H /de{p^\mu_i}(\xi, z^A, y^i , u_i X^\mu + A^\mu_i) X^\mu (\xi,z^A)
\end{equation}
\begin{equation}
\d u_i /d\xi = - \de A^\mu_i/de{x^\mu} (\xi,z^A) - \de H/de{y^i} (\xi, z^A, y^i , u_i X^\mu + A^\mu_i )
\end{equation}
\begin{equation}
\d u /d\xi = - \de H/de{p^\mu_i} (\xi, z^A, y^i , u_i X^\mu + A^\mu_i ) X^\mu(\xi,z^A) u_i 
\end{equation}
\end{subequations}
and represents a system of $2n+1$ ordinary differential equations in the unknowns $y^i(\xi,z^A),u(\xi,z^A),u_i(\xi,z^A)$. \\
First of all, we notice that eqs.(\ref{HJ bicaratteristiche}a), (\ref{HJ bicaratteristiche}b) are independent on $u$, so that eq.(\ref{HJ bicaratteristiche}c) can be integrated after solving the others. \\  
Moreover, the parametric dependence of (\ref{HJ bicaratteristiche}a) and (\ref{HJ bicaratteristiche}b)  on $z^A$ makes their solution depend on $2n$ arbitrary integration functions of the parameters $z^A$, that can be eventually fixed assigning the initial data at $\xi=0$ for the functions $y^i(0,z^A)$ and $u_i(0,z^A) $. 

The fact that the initial submanifold $N$ is represented by the slice $\xi=0$, allows to set the integration functions from the knowledge of the initial data of the field equations \eqref{Dati iniziali}. A first step is performed assigning $y^i(0,z^A) = \psi^i(z^A)$; the second is based on the application of the Legendre transformation, by means of which the normal derivatives of the fields at $N$ are related to the values $u_i(0,z^A)$ by
\begin{equation}\label{imposizione dati iniziali}
\de y^i/de{x^\mu} \bigg|_{\xi=0} n^\mu = \hat\psi^i(z^A) = \de H /de{p^\mu_i}(\xi,z^A,\psi^i(z^A),u_i(0,z^A)X^\mu+A^\mu_i) n^\mu \quad .
\end{equation}
By Dini's theorem, the possibility of calculating $u_i(0,z^A)$ is guaranteed by the requirement
\begin{equation}\label{regularity X}
\det \left(\SD H /de{p^\mu_i}/de{p^\nu_j} X^\mu\,n^\nu\right) \neq 0 \quad .
\end{equation}
The latter is not a direct consequence of the regularity condition \eqref{regularity} on the Hamiltonian function. Whenever the initial submanifold $N$ is fixed, it should be thought as a restriction on the possible choices of the field $X$. 

The result of the above procedure consists in a candidate solution $y^i = y^i(\xi,z^A)$ fulfilling the initial data \eqref{Dati iniziali} and still depending on the choice of the arbitrary functions $A^\mu_i$. 
However, they are not solutions of the whole set of field equations, unless the embeddability conditions \eqref{embeddability} are satisfied. Taking eq.(\ref{HJ bicaratteristiche}a) into account,we notice that the subset of embeddability conditions
\[
\d y^i /d{x^\mu} X^\mu = \de H/de{p^\mu_i} X^\mu 
\] 
is automatically fulfilled. Then, the embeddability conditions only consist in the system of equations
\begin{equation}\label{embeddability 2}
\d y^i /d{z^A} = \de H /de{p^\mu_i}(\xi, z^A, y^i , u_i X^\mu + A^\mu_i) \de x^\mu /de{z^A} 
\end{equation}
that must be fulfilled by the functions $y^i(\xi,z^A)$ and $u_i(\xi,z^A)$. 
The strategy of solution takes advantage of the complete arbitrariness of the functions $A^\mu_i$ and consists in  imposing eqs.\eqref{embeddability 2} on them. 

For this purpose, let us decompose the fields $A_i$ along the basis provided by the adapted coordinates as
\[
A_i = \hat A_i \de /de{\xi} + k^B_i \de/de{z^A} = \left(\hat A_i \de x^\mu /de\xi + k^B_i \de x^\mu /de{z^A} \right) \de /de{x^\mu}
\]
and let us write the embeddability condition in the form
\begin{equation}\label{embeddability decomposed}
\d y^i /d{z^A} = \de H /de{p^\mu_i}\left(\xi, z^A, y^i , (u_i + \hat A_i) X^\mu + k^B_i \de x^\mu /de {z^B}\right) \de x^\mu /de{z^A} \quad .
\end{equation}
Whenever the regularity condition 
\begin{equation}\label{regularity surface}
\det \left(\SD H /de {p^\mu_i}/de{p^\nu_j} \de x^\mu/de{z^A} \de x^\nu/de{z^B} \right) \neq 0 
\end{equation}
 is fulfilled, Dini's theorem allows to locally invert eq.~\eqref{embeddability decomposed} for the functions $k^A_i$, namely
\begin{equation}\label{A}
k^A_i = k^A_i\left(\xi, z^A, y^i, \d y^i /d {z^A}, u_i ,\hat A_i\right) \quad .
\end{equation}
Substituting into \eqref{HJ bicaratteristiche} we obtain:
\begin{subequations}\label{bicharacteristics equations}
\begin{align}
&\De y^i /de{\xi} = f^i\left(\xi, z^A, y^i, \de y^i /de {z^A}, u_i ,\hat A_i\right) \\
&\De u_i /de{\xi} = g_i (\xi,z^A,y^i,u_i,\de y^i/de{z^A}, \de u_i /de{z^A}, \SD y^i /de {z^A} /de{z^B},\hat A_i) 
\end{align}
\end{subequations}
where 
\begin{subequations}\label{g}
\begin{equation}
f^i = \de H /de{p^\mu_i}\left(\xi, z^A, y^i, \de y^i /de {z^A}, u_i\right) X^\mu
\end{equation}
\begin{equation}
g_i = -\de H /de{y^i} \left(\xi, z^A, y^i, \de y^i /de {z^A}, u_i \right) + \de \hat A_i /de\xi + \de k^A_i/de{z^A} (\xi, z^A, y^i, \de y^i /de {z^A}, u_i , \de u_i /de{z^A} , \SD y^i /de{z^A}/de{z^B})
\end{equation}
\end{subequations}
Eqs.\eqref{bicharacteristics equations} represent a system of partial differential equations for the unknowns $y^i$ and $u_i$, in normal form relative to the derivatives with respect to $\xi$. 

We will now show that, for every choice of the function $\hat A_i$, they possess a local solution around the initial manifold corresponding to $\xi=0$, represented by $N$. For this purpose we need to restrict the possible choices  of the initial submanifold using the following regularity assumption. 
\begin{Definition}[Regular initial pair]
A pair $(N,X)$, where $N$ is a submanifold of $M$ having co-dimension $1$ and $X$ is a vector field of $M$ transverse to $N$, is said to be a { regular initial pair} for the Hamiltonian $H$ if satisfies the regularity conditions \eqref{regularity X} and \eqref{regularity surface}, together with 
\begin{equation}\label{solution condition}
\det \left(\SD H /de{p^\mu_i}/de{p^\nu_j} X^\mu X^\nu - \SD H /de{p^\mu_i}/de{p^\nu_j}  \SD H /de{p^\rho_r}/de{p^\lambda_s} X^\mu X^\lambda \de x^\nu/de{z^A} \de x^\rho/de{z^B} \Lambda_{ir}^{AB}\right) \neq 0 \quad , 
\end{equation}
where 
\[
\Lambda^{AB}_{ij} := \left(\SD H /de {p^\mu_i}/de{p^\nu_j} \de x^\mu/de{z^A} \de x^\nu/de{z^B} \right)^{-1}  \quad . 
\]
\end{Definition}
We notice that the conditions defining the regular initial pairs are completely independent of the regularity condition on the Hamiltonian function but have to be considered as restrictions on the possible choices of the initial submanifold $N$ and of the transverse vector field $X$. The hessian on the Hamiltonian plays the role of a metric tensor. \\[2pt]
We are now able to prove the following
\begin{Theorem}
Consider an analytic Hamiltonian section ${\cal S}: p + H(x^\mu,y^i,p^\mu_i)=0$ and a regular initial pair $(N,X)$. Let ${\cal B}\subset {\cal S}$ represent the initial data of the field equations (compare with \eqref{Superficie dati iniziali}) and suppose that $X$ is analytic and satisfies the condition $\de X^\mu/de{x^\mu}=0$. Then, the Hamilton-Jacobi equation of $H$ possesses a family of solutions of  the form 
\begin{equation}\label{sost}
S^\mu(x^\mu,y^i) = \varphi(x^\mu,y^i)\, X^\mu+A^\mu_i(x^\mu)\,y^i \quad , 
\end{equation}
where $A^\mu_i$ are arbitrary functions on $M$. Moreover, there exists a choice of the functions $A^\mu_i$ such that the submanifold ${\cal E}\subset \Lambda^n_2(E)$, locally represented as
\[
{\cal E} :\left\{
\begin{split}
 &y^i = y^i(\xi,z^A) \\  
 &p^\mu_i(\xi,z^A)= u_i (\xi,z^A) X^\mu + A^\mu_i(\xi,z^A) \\
  &p = - H(\xi,z^A,y^i,p^\mu_i)
\end{split}
\right. \quad ,
\]
obtained dragging the initial the submanifold $B$ along the bicharacteristic curves \eqref{bicharacteristics equations} of the Hamilton-Jacobi equation, is a solution of the field equations for every given choice of the initial data.
\end{Theorem}
\begin{Proof}
We already showed that, substituting eq.\eqref{sost} into \eqref{HJ ridotto} and writing the equations for the bicharacteristics, we always determine a set of solutions of \eqref{HJ bicaratteristiche} matching all the possible initial data, and depending parametrically on $A^\mu_i$. \\
Being the regularity condition \eqref{regularity surface} fulfilled, the parameters $k^A_i$ can be (locally) obtained from \eqref{embeddability decomposed} by Dini's theorem, leaving us with a system of partial differential equations \eqref{bicharacteristics equations} in normal form.
Let us now define the functions 
\begin{equation}\label{F}
F^i(\xi,z^A,y^i,u_i,y^i_A,y^i_0) := y^i_0 - f^i\left(\xi, z^A, y^i, y^i_A, u_i \right) \quad , 
\end{equation}
determined from (\ref{bicharacteristics equations}a) choosing an arbitrary analytic function $\hat A_i$. 
Applying Dini's theorem to \eqref{F} it is possible to locally write the surface $F^i = 0$ in parametric form with respect to $u_i$ whenever the condition $\det(\de f^i /de{u_j}) \neq 0$ is fulfilled. 
 Taking eq.~\eqref{embeddability 2} into account, it is easy to calculate
\begin{equation}\label{fu}
 \left( \de f^i/de{u_j}\right) = \left(\SD H /de{p^\mu_i}/de{p^\nu_j} X^\mu X^\nu - \SD H /de{p^\mu_i}/de{p^\nu_j}  \SD H /de{p^\rho_r}/de{p^\lambda_s} X^\mu X^\lambda \de x^\nu/de{z^A} \de x^\rho/de{z^B} \Lambda_{ir}^{AB}\right) \quad . 
\end{equation}
The non singularity of \eqref{fu} is guaranteed by  eq.\eqref{solution condition}.
We can therefore write eq.~\eqref{F} as 
\begin{equation}\label{U}
u_i = U_i(\xi,z^A,y^i,\de y^i/de{z^A},\de y^i/de\xi) \quad . 
\end{equation}
We can also calculate the  derivatives of \eqref{U} as:
\begin{subequations}\label{derivatives}
\begin{equation}
\de u_i /de{\xi} = \de U_i /de\xi + \de U_i /de{y^j} \de y^j/de\xi + \de U_i /de{y^j_A} \SD y^j/de z^A /de\xi + \de U_i /de{y^j_0} \SD y^j /de /de {\xi}
\end{equation}
\begin{equation}
\de u_i /de{z^A} = \de U_i /de{z^A} + \de U_i /de{y^j} \de y^j/de{z^A}+ \de U_i /de{y^j_B} \SD y^j/de z^B /de{z^A} + \de U_i /de{y^j_0} \SD y^j /de \xi /de {z^A} \quad . 
\end{equation}
\end{subequations}
Substituting into eq.(\ref{bicharacteristics equations}b) we obtain:
\[
\de U_i /de\xi + \de U_i /de{y^j} \de y^j/de\xi + \de U_i /de{y^j_A} \SD y^j/de z^A /de\xi + \de U_i /de{y^j_0} \SD y^j /de /de {\xi} = h_i\left(\xi,z^A,y^i,\de y^i/de{z^A},\SD y^i /de \xi /de {z^A},\SD y^i /de z^B /de {z^A}\right) 
\]
being $h_i$ the right hand side of (\ref{bicharacteristics equations}b), after the substitutions of $u_i$ and its derivatives from \eqref{U} and \eqref{derivatives}. 
This equation can be put into normal form with respect to $ \SD y^i /de /de {\xi}$ whenever 
\[
\det \left(\de U_i /de{y^j_0}\right) \neq 0 \quad . 
\]
This condition can be evaluated applying Dini's theorem to eq.~(\ref{bicharacteristics equations}a):
\[
\de U_i /de{y^j_0} = - \left(\de F^i /de {u_k}\right)^{-1} \de F^k /de{y^j_0} =\left( \de f^i/de{u_k}\right)^{-1} \delta^k_j =  \left( \de f^i/de{u_j}\right)^{-1}\quad .
\]
It is non singular by \eqref{solution condition} and \eqref{fu}. This results into a system of second order partial differential equations in Cauchy-Kovalevskaya form as 
\[
\SD y^i /de /de {\xi} = \rho_i \left(\xi,z^A,y^i,\de y^i/de{z^A},\SD y^i /de \xi /de {z^A},\SD y^i /de z^B /de {z^A}\right) \quad . 
\]
In the present case, this means that the system is in normal form for the second order derivatives with respect to $\xi$ and all other derivatives in any variable on the right hand side are of first or second order. Obviously, no second order derivative with respect to $\xi$ appears on the right hand side. Then, by a generalized version of Cauchy-Covalevskaya theorem \cite{friedman},  whenever all the functions defining the system of partial differential equations are analytic, the latter possesses a local solution around the initial surface $\xi=0$, which  represents the initial submanifold $N$. 
\end{Proof}
Now, taking eqs. \eqref{HJ bicaratteristiche} into account,  the general solution of the bicharacteristic equations \eqref{bicharacteristics equations} can be written as 
\begin{equation}\label{solution}
y^i = y^i(\xi,z^A,\alpha_i(z),\beta^i(z)) \quad , \quad u_i = u_i (\xi,z^A,\alpha_i(z),\beta^i(z)) \quad , 
\end{equation}
where the arbitrary functions $\alpha_i(z^A)$ and $\beta^i(z^A)$ can be calculated from the knowledge of the initial data of the field equations on $N$, using the same procedure reported in eq.~\eqref{imposizione dati iniziali}. 

\section{Further developments}

The results obtained up to now prove the existence of a solution of the Hamilton-Jacobi equation for every regular choice of the initial data of the field equations, defined on a regular initial pair. From a computational viewpoint, the advantage of using the bicharacteristic equations over the usual field equation is not great. In this section we will show some ideas that can lead to a technique for solving the Hamilton-Jacobi equation using a direct approach.

The argument uses the technique of the complete integral and is based on the fact that  eq.\eqref{HJ ridotto} is an ordinary Hamilton-Jacobi equation for the function $\varphi(\xi,y^i)$, depending parametrically on the functions $A^\mu_i$. For this reason, it possesses a complete integral of the form
\begin{equation}\label{integrale completo}
\varphi = \varphi(\xi,z^A,y^i,\alpha_i,A^\mu_i)
\end{equation}
satisfying the condition of essential dependence on the parameters 
\begin{equation}\label{essential}
\det \left(\SD \varphi /de y^i /de{\alpha_j} \right) \neq 0  \quad . 
\end{equation}
The parameters $\alpha_i$ result to be first integrals of the problem, i.e. they are constant along the bicharacteristic curves. 
Taking eqs.\eqref{HJ bicaratteristiche} into account, it is easy to prove the following
\begin{Proposition}
The functions $\beta^i = \beta^i(\xi,z^A)$ defined as
\begin{equation}\label{beta}
\beta^i = \de \varphi/de{\alpha_i} (\xi,z^A,y^i,\alpha_i,A^\mu_i)
\end{equation}
are constant along the bicharacteristic curves.
\end{Proposition} 
\begin{Proof}
Taking the derivative along the bicharacteristic curves we obtain that
\[
\d \beta^i /d\xi = \de /de\xi \left(\de \varphi/de{\alpha_i} \right) + \d y^j /d\xi \SD \varphi/de y^j /de{\alpha_i} \quad . 
\]
Using Schwarz theorem and substituting from \eqref{HJ ridotto} we obtain that
\[
\d \beta^i /d\xi = \de /de{\alpha_i}(-H - \de A^\mu_j /de{x^\mu}y^j) + \d y^j /d\xi \SD \varphi/de y^j /de{\alpha^i}=0 \quad . 
\]
The result comes from the fact that $A^\mu_i$ does not depend on $\alpha_i$ and 
\[
\de /de{\alpha^i}(-H) = - \de H/de{p^\mu_i} \SD \varphi /de y^i /de{\alpha_i} X^\mu = - \SD \varphi /de y^i /de{\alpha_i} \d y^i /d\xi 
\]
by eq. (\ref{HJ bicaratteristiche}a). 
\end{Proof}
As a consequence,  the functions $\alpha_i$ and $\beta^1$ can be determined from the knowledge of the initial data. The essentiality condition \eqref{essential} allows to local invert eq.\eqref{beta} with respect to the $y^i$, thus obtaining the candidate solutions of the field equations in the form:
\begin{equation}\label{campi}
y^i = y^i (\xi,z^A,\alpha_i,\beta^i,A^\mu_i)\quad . 
\end{equation}
Eq.\eqref{campi} represents an actual solution of the field equations if it satisfies the embeddability conditions. Taking the derivatives of \eqref{beta} by $z^A$, and substituting the embeddability conditions \eqref{embeddability}, we have that:
\begin{equation}\label{embeddability 3}
\de \beta^i /de{z^A} = \SD \varphi /de{\alpha_i} /de{y^j} \de H /de{p^\mu_j} \de x^\mu /de{z^A} +  \SD \varphi /de{\alpha_i} /de{\alpha_j} \de \alpha_j /de{z^A} +  \SD \varphi /de{\alpha_i} /de{z^A} \quad . 
\end{equation}
By Theorem 2, whenever the initial data take their values on a regular initial pair, there exists a choice of the function $A^\mu_i$ such that the functions \eqref{campi} are a solution of the field equations. However, solving eq.\eqref{embeddability 3} is as complicated as solving the field equations, making the whole argument still partial.
Any further development of this research topic should therefore provide an algorithm including the embeddability condition within the solution technique of the Hamilton-Jacobi equation; this argument is still under analysis. 

\section{An example}

Let us consider the free scalar field in a 4-dimensional Minkowki spacetime $M$, endowed with a metric $\eta = diag(1,1,1,-1)$. It Hamiltonian is given by
\[
H(x^\mu,y,p^\mu) = \frac12 \eta_{\mu\nu} p^\mu p^\nu + \frac12 \mu^2 y^2 \quad . 
\]
Let us consider the problem of determining its evolution starting at $N:x^4=0$ in the positive time direction. We choose $X = \de /de{x^4}$, so that $\xi = x^4$. The surfaces $\xi = const$ are spanned by the coordinates $z^A = x^A, A=1..3)$ and are endowed with a positive definite metric $\delta_{AB}$. 
The regularity of the pair $(N,X)$ can be proved observing that 
\[
\SD H/de{p^\mu}/de{p^\nu} = \eta_{\mu\nu} \quad , 
\]
so that the conditions \eqref{regularity X}, \eqref{regularity surface} and \eqref{solution condition} can be respectively written as:
\[
\eta_{\mu4} X^\mu \neq 0 \quad , \quad \det(\eta_{\mu\nu}\delta^\mu_A\delta^\nu_B )\neq 0 \quad , \quad \eta_{\mu\nu} X^\mu X^\nu \neq 0 \quad . 
\]
This shows that a solution can be obtained whenever $X$ is a timelike vector field and the pullback of the metric tensor on the initial surface is non singular. \\[2pt]
The Hamilton-Jacobi equation becomes
\[
\de S^\mu /de{x^\mu} + \frac12 \eta_{\mu\nu}\de S^\mu /de y \de S^\nu /de y + \frac12 \mu^2 y^2 = 0 \quad . 
\]
Substituting \eqref{sost} we have that
\[
\de \varphi/de\xi + \de A^\mu /de{x^\mu} y + \frac12 \eta_{\mu\nu} X^\mu X^\nu \left( \de \varphi/de y \right)^2 + \frac12  \eta_{\mu\nu} A^\mu A^\nu + \eta_{\mu\nu} X^\mu \de \varphi/de y A^\nu + \frac12 \mu^2 y^2 = 0 \quad . 
\]
We notice that $\eta_{\mu\nu} X^\mu X^\nu = -1 $ and, choosing $\hat A = A^4 = 0$, we also have that $\eta_{\mu\nu} X^\mu A^\nu = 0$.  Substituting $h^A := A^A, A=1..3$ and $\|h\|^2 := \eta_{\mu\nu} A^\mu A^\nu = \delta_{AB} h^A h^B > 0$, we have that
\begin{equation}\label{HJscalar}
\de \varphi/de\xi + \de h^A /de{z^A} y - \frac12 \left( \de \varphi/de y \right)^2 + \frac12  \|h\|^2  + \frac12 \mu^2 y^2 = 0 \quad . 
\end{equation}
Writing $u_1 \sim \de \varphi/de y$, the equations of the bicharacteristics are
\[
\de y /de\xi = - u_1 \quad ; \quad \de u_1/de \xi = -\mu^2 y - \de h^A /de{z^A} \quad . 
\]
Its separation easily leads to 
\[
\de y /de\xi - \SD y /de{z^A}/de{z^B} \delta ^{AB} = \mu^2 y^2 \quad ,
\]
that are identical to the field equations for the free scalar field. 

In order to show what difficulties rise in the application of the arguments of Section IV, 
a complete integral will be determined using the following strategy. Let us consider a polynomial solution
\[
\varphi = \frac12 \mu y^2 + a(\xi,z^A) y + b(\xi,z^A) \quad . 
\]
Substituting into \eqref{HJscalar} we have that
\[
\de a/de\xi y + \de b /de\xi - \frac12 a^2 - a\mu y + \frac12 \|h\|^2 + \de h^A /de{z^A} y = 0 \quad . 
\]
Separating the linear and the homogeneous term in y we can solve for $a(\xi,z^A)$ and $b(\xi,z^A)$ as
\[
a(\xi,z^A) = \alpha(z^A) e^{\mu\xi} + e^{\mu\xi} \int e^{-\mu\xi} \de h^A/de{z^A} d\xi \quad , 
\]
\[
b(\xi,z^A) = \frac12 \int a^2 d\xi + \frac12 \int \| h\|^2 d\xi \quad , 
\]
where the integration constant $\alpha(z^A)$ is the required parameter for the complete integral. We notice that the dependence on $\alpha$ is essential, since
\[
\SD \varphi /de y /de\alpha = e^{\mu\xi} \neq 0 \quad . 
\]
Then we can calculate
\[
\beta = \de\varphi/de\alpha = e^{\mu\xi} y + \int a\, e^{\mu\xi} d\xi \quad,
\]
whence
\begin{equation}\label{soluzione scalare}
y(\xi,z) =  e^{-\mu\xi} \beta -  e^{-\mu\xi} \int a\, e^{\mu\xi} d\xi \quad . 
\end{equation}
Eq.\eqref{soluzione scalare} represents the candidate solution of the problem. It also has to fulfill the embeddability conditions, that can be calculated as follows:
\[
\de \beta /de{z^A} = e^{\mu\xi} h_A + \frac{e^{2\mu\xi}}{2\mu} \de \alpha /de{z^A} + e^{2\mu\xi} \int e^{-\mu\xi} \SD h_B /de z^B /de{z^A} d\xi \, d\xi \quad . 
\]
The difficulties of this last equation , representing the embeddability conditions, are similar to those of solving the field equations directly. In fact, it can be easily written in the form
\[
\SD h_B /de z^B /de{z^A} - \SD h_A /de /de\xi = \mu^2 h_A \quad ,
\]
whence the solution can be easily found.

\end{document}